\documentstyle[12pt]{article}
\begin{document}
\title{Interplay of real space and momentum space topologies
 in strongly correlated fermionic systems.}
\author{G.E. Volovik\\
Low Temperature Laboratory,\\ Helsinki
  University of Technology\\
 Box 2200, FIN-02015 HUT, Finland\\
and\\
L.D. Landau Institute for Theoretical Physics, \\
117334 Moscow,
  Russia}

\date{\today}
\maketitle
\begin{abstract}
 We discuss the momentum-space topology of 3+1 and 2+1 strongly correlated
fermionic systems. For the 3+1 systems the important universality
class is determined by the topologically stable Fermi points in
momentum space. In the extreme limit of low energy the condensed matter
system of this universality class acquires all the symmetries, which we
know today in high energy physics: Lorentz invariance, gauge invariance,
general covariance, etc.  The chiral fermions as well as gauge bosons and
gravity field arise as fermionic and bosonic collective modes of the
system. This introduces the conceptual similarity between such condensed
matter and the quantum vacuum of Standard Model, which also contains the
Fermi point. For the 2+1 fermionic systems the momentum space topology
gives rise to the quantization of different physical parameters, such as
Hall conducivity and spin-Hall conductivity, and to quantum phase
transitions with the abrupt change of the momentum-space topology. The
combined ${\bf p}$-space and ${\bf r}$-space topology determines in
particular the topology of the energy spectrum of fermion zero modes
living within the topological defects; fermionic charges of the
topologically nontrivial extended objects; axial anomaly; etc.
\end{abstract}

\tableofcontents

\section {Introduction}

As a rule, in strongly correlated fermionic systems, there are no small
parameters which allow us to treat them perturbatively. Also
there are not so many models which can be solved exactly. That is why the
qualitative description, which is based on the universal features coming
from the symmetry and topology of the ground state, is instructive. In
particular, it allows us to construct the effective low-energy theory of
strongly correlated fermionic systems of given universality class, which
incorporates all the important features of this class. All the
information on the symmetry and topology of the strongly correlated
fermionic system is contained in the low-energy asymptote of the
Green's function of the fermionic fields, which is characterized by
the topological quantum numbers (See the book \cite{ThoulessBook} for
review on the role of the topological quantum numbers in physics).

In a semiclassical aproximation
the Green's function of the fermionic field propagating in inhomogeneous
environment depends on 4-momentum and on coordinates in 3+1 space-time:
$G(p_0,{\bf p}, t, {\bf r})$. If the fermionic field has different
species characterized by spin, isospin and other fermionic charges, the
Green's function becomes the complicated object with nontrivial topology
in combined 4+3+1 space (here and further we assume that the frequency
$p_0$ is on the imaginary axis to avoid singularities of the Green's
function on the mass shell).  The topology in this combined space
includes as particular cases: (i) the topologically  nontrivial
configurations in conventional ${\bf r}$-space, such as vortices
(strings), monopoles, textures, solitons, etc., and (ii) the
${\bf p}$-space topology of the homogeneous fermionic vacua, which
determines the universality classes of the low-energy quasiparticle
spectrum (Sec. \ref{MomentumSpaceTopology}) and some classes of the
quantum phase transitions (Sec. \ref{QuantumPhaseTransition}). In
addition there can be  nontrivial interconnections between the space-time
and the energy-momentum space topologies  (see {\it  e.g.}
Refs.\cite{VolovikMineev1982} and
\cite{Grinevich1988}). This interplay between the ${\bf r}$-space and
${\bf p}$-space topologies determines in particular:  (iii) the topology
of the energy spectrum of fermion zero modes living within the
topological defects (Sec. \ref{CombinedTopologySspectrumVortex}); (iv)
quantization of physical parameters (Sec.
\ref{InterplayAndQuantization});  (v) fermionic charges of the
topologically nontrivial extended objects (Sec.
\ref{InterplayAndQuantization}); (vi) axial anomaly (Sec.
\ref{InterplayTopologiesChiralAnomaly}); (vii) edge states, etc.

\section {Momentum space topology of Fermi point.}
\label{MomentumSpaceTopology}

Momentum space topology of the Green's function determines the
manifold of zeroes of the quasiparticle energy spectrum. We
consider here the universality class of fermionic systems, which have
Fermi points -- the points in the 3D  ${\bf p}$-space where the energy
$E({\bf p})$ of fermionic quasiparticle is inevitably zero. Standard Model
and superfluid $^3$He-A and representatives of this universality class.
In strongly correlated fermionic systems of this class the chirality, Weyl
fermions, gauge fields and gravity emergently appear in the low
energy corner, together with the corresponding symmetries which include
the Lorentz symmetry and local $SU(N)$ symmetry. This phenomenin,
following solely from the topology (and symmetry) of the Fermi point
supports the ``anti-grand-unification'' scenario according to which
Standard Model (even in its GUT extension), is an effective theory, which
is applicable only in the infrared limit.  Most of the symmetries of this
effective theory are the attributes of the theory: the symmetries
gradually appear in the low-energy corner together with the effective
theory itself.

\subsection{Topological invariant for Fermi point}

 In the most simple case the stability of the Fermi point is
provided by the following topological invariant in terms of the matrix
Green's function ${\cal G}({\bf p},p_0)$:
\begin{equation}
N_3 = {1\over{24\pi^2}}e_{\mu\nu\lambda\gamma}~
{\bf tr}\int_{\sigma}~  dS^{\gamma}
~ {\cal G}\partial_{p_\mu} {\cal G}^{-1}
{\cal G}\partial_{p_\nu} {\cal G}^{-1} {\cal G}\partial_{p_\lambda}  {\cal
G}^{-1}~.
\label{TopInvariant}
\end{equation}
The integration here is around  the 3-dimensional surface
$\sigma$ embracing the Fermi point at ${\bf p}={\bf p}_F$, $p_0=0$ in
the 4D space $({\bf p},p_0)$. It
can be easily checked that under continuous variation of the matrix
function ${\cal G}({\bf p},p_0)$ the integrand changes by a total
derivative. That is why the integral over the closed 3-surface does not
change, i.e. $N_3$ is invariant under continuous deformations of the
Green's function ${\cal G}({\bf p},p_0)$, and also it is independent of
the choice of closed 3-surface around the Fermi point.
This the generalization of the momentum space
topology of chiral fermions introduced in Ref.
\cite{NielsenNinomiya}.

The possible
values of the invariant can be easily found: if one chooses the matrix
function which changes in
$U(2)$ space one obtains the integer values of $N_3$. They describe the
mapping  of the $S^3$ sphere surrounding the singular point in 4-space of the
energy-momentum
$(p_0,{\bf p})$ into the $SU(2)=S^3$ space of the Green's function.  The
same integer values
$N_3$ are preserved for any Green's function matrix, if it is well
determined outside the singularity where ${\rm det} ~G^{-1}\neq 0$.  The
index $N_3$ thus represents topologically different Fermi points -- the
singular points in the 4D space $({\bf p},p_0)$.

\subsection{Topological invariant as the generalization of chirality.}
\label{GeneralizationChirality}

An example of the Fermi points with $N_3=\pm 1$ is provided by
massless neutrino -- the massless spin-1/2 particle whose $2\times 2$
matrix Green's function and  Hamiltonian are
\begin{equation}
{\cal G} =(ip_0-{\cal H})^{-1}~,~ {\cal H}=C
c{\bf \sigma}\cdot {\bf p}~,
\label{Neutrino}
\end{equation}
where $c$ is the speed of light; ${\bf \sigma}$ are the Pauli spin
matrices; $C$ is the chirality, with
$C=+1$ or
$C=-$ for the right-handed and left-handed neutrino: the spin of the
particle is oriented along or opposite to its momentum, respectively. The
Green's function is determined on the imaginary  frequency axis,
$z=ip_0$. That is why  this propagator has no singularity  on the mass
shell, and the only singularity is at the point $(p_0=0,{\bf p}=0)$ in the
4D momentum-frequency space.

The energy spectrum of neutrino,  $E({\bf
p})=cp$, becomes zero at ${\bf p}=0$. From the consideration of the
energy spectrum it is not clear whether this zero survives if the
interaction is introduced, or if the Lorentz symmetry is violated and the
spectrum becomes non-linear. The mass protection of neutrino is provided
by the non-zero value of the topological invariant
$N_3$ in Eq.(\ref{TopInvariant}) , whose value depends on the chirality of
neutrino
$N_3=C$. Since any perturbation of the Green's function cannot change
the value of the topological invariant, the point singularity of the
Green's function in $(p_0,{\bf p})$-space is robust to interaction and
violation of relativistic invariance. The location of the singularity can
be shifted but the singularity cannot disappear.  If the vacuum has
$N_3\neq 0$, some of its fermionic excitations must be massless
(gapless) because of the topological stability of the Fermi point.

For a single chiral fermion in Eq.(\ref{Neutrino}) the meaning
of this topological invariant can be easily visualized. Let us consider
the behavior of the particle spin ${\bf s}({\bf p})$ as a function the
particle momentum ${\bf p}$ in the 3D-space  ${\bf p}=(p_x,p_y,p_z)$. For
the right-handed particle, whose spin is parallel to the momentum, one has
${\bf s}({\bf p})={\bf p}/2p$, while for left-handed ones ${\bf s}({\bf
p})=-  {\bf
p}/2p$. In both cases the spin distribution in the momentum space looks
like a hedgehog, whose spines are represented by spins. Spines point
outward for the right-handed particle producing the mapping of the sphere
$S^2$  in 3D momentum space onto the sphere $S^2$ of the spins with
the index $N_3=+1$. For the left-handed particle the spines of the
hedgehog look inward and one has the mapping with $N_3=-1$.  In the
3D ${\bf p}$-space such hedgehogs are topologically stable and thus
$E^2({\bf p})$ must be inevitably zero somewhere.

\subsection{Relativistic massless chiral fermions emerging near Fermi
point.}
\label{RelativisticMassless}

The most remarkable property of any, even non-relativistic, system, which
has a topologically stable Fermi point with  $N_3=+ 1$ or $N_3=-1$, is
that the relativistic Eq.(\ref{Neutrino}) is always reproduced in the
low-energy corner.  Let us consider the Green's function in the 3+1
``Minkowskii'' momentum-energy space
$p_\mu=(p_0,{\bf p})$, where the frequency $p_0$ is now on the real
axis. If $p_{\mu}^{(0)}$ is the position of the Fermi point, the
expansion of the inverse propagator in terms of the deviations from this
Fermi point, $p_\mu -p_{\mu}^{(0)}$ gives
\begin{equation}
{\cal G}^{-1}=\sigma^a e^\mu_a (p_{\mu} - p_{\mu}^{(0)})
~.
\label{GeneralPropagator}
\end{equation}
Here $\sigma^a=(1,{\bf \sigma})$, and $e^\mu_a$ is the matrix in the
linear expansion of the inverse propagator, which plays the part of the
effective vierbein. The quasiparticle spectrum
$E({\bf p})$ is given by the poles of the propagator, and thus by equation
\begin{equation}
g^{\mu\nu}(p_{\mu} - p_{\mu}^{(0)})(p_{\nu} -
p_{\nu}^{(0)})=0~,~g^{\mu\nu}=\eta^{ab}e^\mu_a
e^\nu_b ~,
\label{GeneralEnergy}
\end{equation}
where $\eta^{ab}=diag(-1,1,1,1)$. Thus in the vicinity of the Fermi point
the quasiparticles are chiral massless fermions propagating in the
background chracterized by the Lorentzian metric
$g^{\mu\nu}$ and by gauge field $A_\mu=p_{\mu}^{(0)}$. Both fields,
$g^{\mu\nu}$ and $A_\mu$, are dynamical bosonic collective modes of the
fermionic vacuum. Thus the fermionic part of the effective low-energy
action (and even some part of the bosonic action, obtained by
integration over the ``relativistic'' fermions, see detailed discussion in
the review paper \cite{PhysRepReview}) acquires all the attributes of the
relativistic quantum field theory.  This implies that the classical (and
also the quantum) gravity and gauge fields are not fundamental, but
inevitably appear together with chiral fermions in the low energy corner
as collective bosonic and fermionic zero modes of quantum vacuum, if this
vacuum belongs to the universality class with Fermi point.

\subsection{Topology protected by symmetry.
Mass protection in Standard Model}
\label{TopologyAndSymmetry}

Typically there are several species of fermionic excitations of
quantum vacuum. In the Standard Model each generation contains 8
left-handed and 8 right-handed fermions. The topological
invariant in Eq.(\ref{TopInvariant}), where the trace is over all the
fermionic indices, is zero for the Fermi point in the Standard Model:
$N_3=0$. Nevertheless the fermions are massless above the electroweak
scale  $\sim$200 GeV. The mass protection for the Standard Model
fermions is provided by the interplay of topology and symmetry. Let us
introduce the matrix
${\cal N }$ whose trace gives the invariant
$N_3$ in Eq.(\ref{TopInvariant}):
\begin{equation}
{\cal N} = {1\over{24\pi^2}}e_{\mu\nu\lambda\gamma}~
 \int_{\sigma}~  dS^{\gamma}
~ {\cal G}\partial_{p_\mu} {\cal G}^{-1}
{\cal G}\partial_{p_\nu} {\cal G}^{-1} {\cal G}\partial_{p_\lambda}  {\cal
G}^{-1}~,
\label{TopInvariantMatrix}
\end{equation}
where as before the integral is about the Fermi point in the 4D
momentum-energy space. Let us consider the expression
\begin{equation}
({\cal N},{\cal P}) = {\bf tr} \left[{\cal N P}\right]~,
\label{NP}
\end{equation}
where ${\cal P}$ is either the discrete symmetry or the generator of
continuous symmetry of the vacuum. Since the vacuum is  ${\cal
P}$-invariant, ${\cal P}$ commutes with the Green's
function, and Eq.(\ref{NP}) is robust to any deformation of the
Green's function, which retains the symmetry
${\cal P}$. The Eq.(\ref{NP}) is the topological
invariant protected by symmetry.

Let us consider the following $Z_2$ symmetry of the  Standard Model:
\begin{equation}
{\cal P}=e^{2\pi i {\cal W}_3^L } ~~,
\label{DiscreteSymmetry}
\end{equation}
where ${\cal W}_3^L$ is the generator of the $SU(2)_L$ weak rotations.
Since the vacuum above the electroweak scale is  $SU(2)$
invariant, the Eq.(\ref{NP}) is the right symmetry protected invariant.
Its value
\begin{equation}
 \left( {\cal P},{\cal N} \right)~
= 16 ~,
\label{16}
\end{equation}
 shows that all 16 fermions of
one generation are massless above the electroweak scale.
Below the electroweak scale, this ${\cal P}$ is no more the symmetry of
the vacuum. Moreover there is no symmetry, which could produce the
non-zero invariant of the type of  Eq.(\ref{NP}). All the generators of
the  remaining symmetry $U(1)_Q\times SU(3)_c$ give zero value, e.g. for
the generator of the $U(1)_Q$ group of the electromagnetism one has
$({\cal N},{\cal Q})=0$. This means that the Fermi point has no
topological stability, all the Green's function singularitise can be
removed by interaction and thus all the fermions are massive below the
electroweak scale.

The same happens in some triplet superfluids and superconductors, e.g.
in the planar state of $^3$He, where $N_3=0$ for the Fermi point, but the
gapless fermions are protected by the proper discrete symmetry ${\cal P}$
of the superfluid vacuum, which gives $({\cal N},{\cal P})=2$
\cite{PhysRepReview}.  It is important that the gap protection
by non-zero topological invariant, $({\cal N},{\cal P})\neq 0$, remains
valid even in the non-relativistic system, such as the planar phase of
$^3$He, where the chirality itself appears only gradually in the
low-energy corner. Applying this to the relativistic systems, such as
Standard Model, one obtains that the topological mass protection in
Eq.(\ref{16}) remains valid even if the Lorentz symmetry is violated at
higher energy of Planck scale, where the Dirac equation is not applicable
any more. The only requirement is that the discrete symmetry ${\cal P}$
still remains at the Planck scale. Otherwise the Planck physics will
provide the intrinsic mass to the fermions, though in the low-energy
corner this mass can be much smaller than all the characteristic energy
scales, since it contains the Planck energy in denominator.

\subsection{Interplay of ${\bf r}$-space and ${\bf p}$-space topologies
and chiral anomaly.}
\label{InterplayTopologiesChiralAnomaly}

The chiral anomaly is the phenomenon which allows the nucleation of the
fermionic charge from the vacuum \cite{Adler,BellJackiw}. Such
nucleation results from the spectral flow of the fermionic charge through
the Fermi point to high energy.  In
relativistic theories the rate of production of some fermionic charge $q$
from the vacuum by applied electric and magnetic fields is
\begin{equation}
\dot q=\partial_\mu J^\mu ={1\over {8\pi^2}} \sum_a C_a q_a e_a^2
F^{\mu\nu}F^{*}_{\mu\nu}~,
\label{ChargeParticlProduction}
\end{equation}
Here $q_a$ is the charge carried by the $a$-th fermion (when the fermions
is nucleated due to axial anomaly, its charge $q_a$ also enters
the world from the vacuum); $e_a$ is the charge of the
$a$-th fermion with respect to the gauge field $F^{\mu\nu}$; $C_a=\pm 1$
is the chirality of the $a$-th fermion; and $F^{*}_{\mu\nu}$ is the dual
field strength.

In a more general case when the chirality is not readily defined, for
example in superfluid $^3$He-A, the above equation can be
presented via the momentum-space topological invariant
\begin{equation}
\dot q=  {1\over {8\pi^2}}
\left( {\cal Q}{\cal E}^2, {\cal N} \right) F^{\mu\nu}F^{*}_{\mu\nu}~,
\label{ChargeParticlProductionGeneral}
\end{equation}
where ${\cal Q}$ is the matrix of the charges $q_a$, whose creation we
consider, and
${\cal E}$ is the matrix of the ``electric'' charges $e_a$.

For example, the rate of baryon production in the Standard Model by
the $SU(2)_L$ weak field takes the form
\begin{equation}
\dot B={ 1\over 4\pi^2}  \left(  ({\cal W}_3^L)^2 {\cal
B},{\cal N} \right) {\bf B}^i_W\cdot {\bf
E}_{iW} ~,
\label{TopologyBaryoProduction}
\end{equation}
where ${\bf B}^i_W$ and ${\bf E}_{iW}$ are the $SU(2)_L$ magnetic and
electric fields.

This is an example of interplay between the  momentum-space
and real-space topologies: the  Eq.(\ref{TopologyBaryoProduction})
represents  the density of the topological charge in $({\bf r},t)$-space,
$(1/4\pi^2){\bf B}^i_W\cdot {\bf E}_{iW}$,  multipled by the factor
$\left(  ({\cal W}_3^L)^2 {\cal B},{\cal N} \right)$,  which is the
topological invariant in the $({\bf p},p_0)$-space.
The nucleation of baryons occurs when the topological charge of the vacuum
changes, say, by sphaleron or due to de-linking of linked loops of the
cosmic strings \cite{tvgf,jgtv,barriola}.

Since the Eq.(\ref{TopologyBaryoProduction}) is completely
determined by the two topologies and does not
require the Lorentz invariance, it can be
applied to $^3$He-A. The dynamical effective gauge field  is
simulated in $^3$He-A by the collective mode of unit vector field
$\hat{\bf l}$, which is responsible for the change of position of the
Fermi point:
${\bf A}= p_F\hat{\bf l}$. The fermionic charge which is measurable in
$^3$He-A is the linear momentum. When the $a$-th quasiparticle is created
at the Fermi point its momentum ${\cal{\bf
P}}_a=-C_a p_F\hat{\bf l}$ is transferred from the inhomogeneous vacuum
to the world of quasiparticles. The rate of the momentum transfer is
\begin{equation}
{\dot{\bf P}}= { 1\over 4\pi^2} \left(  {\cal{\bf P}} {\cal E}^2,{\cal N}
\right)
{\bf B} \cdot {\bf E} = { 1\over 4\pi^2}
{\bf B} \cdot {\bf E} \sum_a  {\bf P}_a C_a e_a^2 ~.
\label{MomentumProductionGeneral}
\end{equation}
Here ${\bf B}=(p_F/\hbar)\nabla\times\hat{\bf l}$ and   ${\bf E}= (p_F/\hbar)
\partial_t
\hat{\bf l}$ are effective ``magnetic''  and ``electric'' fields which
are simulated by the space-dependent and time -dependent texture of the
$\hat{\bf l}$-vector;
${\cal E}$ is the matrix of corresponding ``electric'' charges:
$e_a=-C_a$ (the ``electric'' charge is opposite to the chirality of the
$^3$He-A
quasiparticle). Using this translation to the $^3$He-A language one
obtains that the momentum production from the texture per unit time per unit
volume is
\begin{equation}
 {\dot{\bf P}}=- {p_F^3\over 2\pi^2 \hbar^2}\hat{\bf l}
\left(\partial_t\hat{\bf l}\cdot(\nabla\times \hat{\bf l})\right)
    ~.
\label{MomentumProductionAPhase}
\end{equation}
Integration of the momentum production over the cross section of the
moving $\hat{\bf l}$-texture gives the net force acting on the texture,
which was measured in experiments on rotating $^3$He-A \cite{BevanNature}.
These experiments confirmed the anomalous nucleation of fermionic
charge described by the Adler-Bell-Jackiw equation.

\section{Interplay of topologies in quantization of parameters}
\label{InterplayAndQuantization}
\subsection{Nontrivial
momentum space topology of fully gapped systems}
\label{NontrivialMomentum}

The fully gapped systems have no singularities in the momentum
space, nevertheless the momentum-space topology of their vacua can be
nontrivial.  Such vacuum states with nontrivial  momentum-space
topology have the counterpart in a real space: the extended objects
with no singularity in the order parameter field, whose topology
is neverheless nontrivial. These are textures (or skyrmions)
characterized by the homotopy group $\pi_3$ in 3D space and $\pi_2$ in 2D
space.

Typically the nontrivial momentum-space topology of the gapped
systems is important in 2+1 dimensions,  e.g. for the 2D electron system
exhibiting the quantum Hall effect
\cite{Kohmoto,Ishikawa}; for thin film of
$^3$He-A (see
\cite{VolovikYakovenko}; for the 2+1
superconductors with broken time reversal symmetry
\cite{VolovikEdgeStates}; for the 2+1 world of fermions living within the
domain walls, etc.  The ground states, vacua, in 2+1 systems are
characterized by the momentum space topological invariants, which are
obtained from the dimensional reduction of the invariants for the Fermi
oint. In particular, the dimensional reduction of the topological charge
$N_3$ in Eq.(\ref{TopInvariant}) leads to the following invariant
describing the ground state of 2+1 system:
\begin{equation}
N_3 = {1\over{24\pi^2}}e_{\mu\nu\lambda}~
{\bf tr}\int   d^2pdp_0
~ {\cal G}\partial_{p_\mu} {\cal G}^{-1}
{\cal G}\partial_{p_\nu} {\cal G}^{-1} {\cal G}\partial_{p_\lambda}  {\cal
G}^{-1}~,
\label{3DTopInvariant}
\end{equation}
where the integral now is over the whole 3D $(p_0,p_x,p_y)$-space. (If
the crystalline system is considered the integration over $p_x,p_y$ is
bounded by Brillouin zone.) The integrand  is determined everywhere in
this 3D space since the system is fully gapped and thus the Green's
function is nowhere singular.

In thin films, in addition to spin indices,
the Green's function matrix ${\cal G}$ contains the indices of the
transverse levels, which come from quantization of motion along the
normal to the film \cite{Exotic}. Fermions on different transverse levels
represent different families of fermions with the same properties. This
would correspond to generations of fermions in the Standard Model, if our
3+1 world is situated within the soliton wall in 4+1 space-time.

Dimensional reduction of the topological invariants protected by symmetry
in Eq.(\ref{NP}) leads to other charges characterizing the
ground state of 2+1 systems. These charges were discussed in
Refs. \cite{Yakovenko2} and \cite{Exotic}. Further dimensional
reduction determines the topology of the edge states: the fermion zero
modes, which appear on the surface of the 2D system or within the domain
wall separating domains with different values of $N_3$ (see Ref.
\cite{VolovikEdgeStates}).

\subsection{2+1 systems with nontrivial
${\bf p}$-space topology}
\label{ExamplesNontrivialMomentum}

An example of the 2+1 system with nontrivial $N_3$ is the
crystal layer of the chiral $p$-wave superconductor, where both time
reversal symmetry and reflection symmetry are spontaneously broken.
Current belief holds that  such a superconducting state occurs in the
tetragonal Sr$_2$RuO$_4$ material
\cite{Rice,Ishida}. The  Bogoliubov-Nambu  Hamiltonian
for the 2+1 fermions living in the layer is
\begin{equation}
{\cal H}= {\check \tau}^i g_i({\bf p})~~,~~g_3 ={p_x^2+p_y^2\over
2m^*}-\mu~,~g_1=  c p_x~,~g_2=  c p_y  ~,~ c={\Delta \over p_F}~.
\label{2DFullyGapped}
\end{equation}
Here $m^*$ is the effective mass
of quasiparticles in normal Fermi liquid above the superconducting
transition; $\Delta$ is the gap in superconducting state; ${\check
\tau}^i$ are the Pauli matrices describing the Bogoliubov-Nambu spin
in superconductors; we consider fermions with a given projection of the
ordinary spin. The quasiparticle spectrum $E({\bf p})$ is fully gapped,
since
\begin{equation}
E^2={\cal H}^2={\bf g}^2({\bf p})=\left({p^2\over
2m^*}-\mu\right)^2+c^2p^2~,
\label{SquareEnergy}
\end{equation}
 is nowhere zero
(except for the special case
$\mu=0$ dicussed later in Sec. \ref{QuantumPhaseTransition}, where the
quasiparticle energy is zero at the point
$p_x=p_y=0$).

There
are two extreme cases:
\begin{eqnarray}
m^*c^2\gg |\mu|~:~~~~E^2\approx
\mu^2 +c^2(p_x^2+p_y^2)~;\label{2DNambuVsDirac2}\\
m^*c^2\ll \mu  ~:~~~~E^2 \approx
 v_F^2(p-p_F)^2 +
\Delta^2~.
\label{2DNambuVsDirac}
\end{eqnarray}
In Eq.(\ref{2DNambuVsDirac2}) the  Bogoliubov-Nambu  Hamiltonian
asymptotically approaches  the 2+1 Dirac Hamiltonian in the limit
$m^*c^2 \gg |\mu|$, with minimum of the energy spectrum being
at $p=0$. The Eq.(\ref{2DNambuVsDirac}) corresponds to the real
situation in superconductors where
$\Delta\ll \mu$ and thus $m^*c^2 \ll |\mu|$. In this limit case the
minimum of the energy spectrum is at $p=p_F$:

The topological invariant $N_3$ in Eq.(\ref{3DTopInvariant}) can be
expressed in terms of the vector field ${\bf g}({\bf p})$ in the momentum
space:
\begin{equation}
N_3= {1\over 4\pi}\int {d^2p\over |{\bf g}|^3}~{\bf
g}\cdot \left({\partial {\bf
g}\over\partial {p_x}} \times {\partial {\bf
g}\over\partial {p_y}}\right)~.
\label{2DInvariant}
\end{equation}
Since at infinity the unit vector ${\bf
g}/|{\bf g}|$ approaches the same value, $({\bf g}/|{\bf g}|)_{p \rightarrow
\infty}
\rightarrow (0,0,1)$, the 2D momentum space $(p_x,p_y)$ is isomorhic to the
compact
$S^2$ sphere. That is why the invariant
$N_3$ describes the mapping of this $S^2$ to the $S^2$
sphere of the unit vector ${\bf g}/|{\bf g}|$.  One finds that
$N_3=1$ for
$\mu>0$ (we considered here only one spin projection, the trace over
spins gives $N_3=2$ in Eq.(\ref{3DTopInvariant})), and
$N_3=0$ for
$\mu<0$.

If the time reversal symmetry is broken in a $d$-wave
superconductor, the superconducting state is
fully gapped. The Bogoliubov-Nambu  Hamiltonian
for the 2+1 fermions living in the layer is given by
Eq.(\ref{2DFullyGapped}) with $g_1= d_{x^2-y^2} (p_x^2-p_y^2)$, $g_2=
d_{xy} p_xp_y$, where $d_{x^2-y^2}$ and $d_{xy}$ are the
corresponding components of the order parameter. The topological invariant
in Eq.(\ref{2DInvariant}) per one layer of such $d_{x^2-y^2}+id_{xy}$
state is $N_3=\pm 2$ for $\mu>0$ \cite{VolovikEdgeStates}. (If both spin
components are taken into account one has $N_3=\pm 4$; the sign of
$N_3$ is determined by the  sign of $d_{xy}/d_{x^2-y^2}$.) And again
$N_3=0$ for
$\mu<0$.

\subsection{Quantum phase transition as change of momentum space
topology}
\label{QuantumPhaseTransition}

At $\mu=0$ there is a quantum phase transition between the
two vacuum states, with $N_3\neq 0$ and $N_3=0$. The two states have the
same internal symmetry, but different momentum-space topology. The
intermediate state between these two fully gapped states, which occurs at
$\mu=0$, is gapless: $E_{\mu=0}({\bf p}=0)=0$. In this gapless state the
topological invariant in Eq.(\ref{3DTopInvariant}) is ill-defined, since
the Green's function has singularity.

In a $d$-wave superconductor the quantum phase transition can occur at
$\mu>0$ if the order porameter $d_{xy}$ changes sign. When $d_{xy}$
crosses zero the invariant in Eq.(\ref{2DInvariant}) changes from
$N_3=+2$ to $N_3=-2$. The intermediate state with $d_{xy}=0$ is also
gapless.

The quantum phase transition between the states with different $N_3$ can
be also organized by changing the effective mass $m^*$. The transition
between states  with $N_3\neq 0$ and $N_3=0$ occurs when at fixed $\mu
\neq 0$ the inverse mass $1/m^*$ crosses zero. In the $p$-wave
superconductor the intermediate state at $1/m^*=0$ corresponds to the
Dirac vacuum in 2+1 system, where (see Eq.(\ref{2DNambuVsDirac2}))
$|\mu|\ll  |m^*|c^2$ and $M({\bf p})\rightarrow -\mu$. This
intermediate state is fully gapped everywhere, and the topological
invariant in Eq.(\ref{2DInvariant}) for this gapped Dirac vacuum
is well defined and has the fractional value $N_3=1/2$. All this happens
because the momentum space is not compact in this intermediate state: the
unit vector ${\bf g}/|{\bf g}|$ does not approach the same value at
infinity.

The fractional topology, $N_3=1/2$, of the intermediate Dirac  state
demonstrates the  marginal behavior of the vacuum of 2+1 Dirac fermions.
The physical properties of the vacuum, which are related to the
topological quantum numbers in momentum space (see below), are not well
defined for the Dirac vacuum. They crucially depend on how the Dirac
spectrum is modified at high energy: towards
$N_3=0$ or towards $N_3=1$.

\subsection{Quantization of physical parameters}
\label{QuantizationPhysicalParameters}

The topological invariants of the type in Eq.(\ref{3DTopInvariant}) determine
the anomalous properties of the 2+1 systems. In particular, they are
responsible for quantization of physical parameters, such as Hall conductivity
\cite{Ishikawa} and spin Hall conductivity
\cite{VolovikYakovenko,Senthill,ReadGreen}. The Eq.(\ref{3DTopInvariant})
leads to quantization of the $\theta$-factor \cite{VolovikYakovenko}
\begin{equation}
  \theta ={\pi\over 2}~ N_3
 ~ ,
\label{Theta}
\end{equation}
in front of the Chern-Simons term
\begin{equation}
 S_\theta =  { \hbar \theta\over{32\pi^2}}\int
d^2x\hskip1mm dt\hskip1mm e^{\mu\nu\lambda}
  A_\mu   F_{\nu\lambda} ~,~F_{\nu\lambda}=\partial _\nu A_\lambda -
\partial_\lambda A_\nu=
\hat {\bf d}\cdot\left(\partial_\nu\hat {\bf d}\times\partial_\lambda\hat
{\bf d}\right) ,
\label{ChernSimonsFilm}
\end{equation}
where $\hat {\bf d}$ is unit vector characterizing the fermionic spectrum.
The $\theta$-factor determines the quantum statistics of skyrmions -- the
nonsingular topological objects of the $\hat {\bf d}$ field described by the
homotopy group $\pi_2$. In $^3$He-A film skyrmions are either fermions or
bosons depending on the thickness of the film: when the film grows the
quantum transitions occur successively, at which the momentum-space
invariant $N_3$ and thus the statistics of fermions
abruptly change \cite{Exotic}. These quantum
(Lifshitz) transitions between the states with different $N_3$ occur
through the intermediate gapless regimes where $N_3$ is not well-defined.

The Chern-Simons action in Eq.(\ref{ChernSimonsFilm}) represents the product of
two topological invariants: $N_3$ in $({\bf p},p_0)$-space and
Hopf invariant in 2+1 coordinate space-time. This is an example of
topological term in action characterized by combined
${\bf p}$-space and ${\bf r}$-space topologies.

Another consequence of interplay of topologies in ${\bf p}$- and ${\bf
r}$- spaces is that the topological invariants protected by the
corresponding symmetries in Eq.(\ref{NP}) determine the fermionic charges
of topological objects -- skyrmions. These are spin, electric
charge, etc.,  \cite{Exotic,Yakovenko2}.

\subsection{Quantization of spin Hall conductivity}

Lets us consider the spin quantum Hall effect which can arise, in
particular, in electrically neutral systems, such as $^3$He-A. An external
magnetic field  ${\bf H}(x,y,t)$ interacts with nuclear spins, producing
the Pauli term
${1\over 2}\gamma{\bf
\sigma}
\cdot{\bf H}$ in the Hamiltonian. As a result the Green's function
contains the long time derivative $-i\partial_t -{1\over 2}\gamma{\bf
\sigma}
\cdot{\bf H}$. This corresponds to the introduction of the
${\bf A}_0$ component of the external $SU(2)$ gauge field:
\begin{equation}
{\bf A}_0= \gamma{\bf H}~.
\label{H}
\end{equation}
To calculate the spin current density induced by this field, it is
convenient to introduce the auxiliary components, ${\bf A}_1$ and
${\bf A}_2$, of the $SU(2)$ gauge field.

Therefore it is instructive
to introduce the general external $SU(2)$ gauge field
${\bf A}_\mu$ with
nonzero curvature ${\bf F}_{\mu\nu}$:
\begin{equation}
{\bf F}_{\mu\nu} = \partial_\mu{\bf A}_\nu  -
\partial_\nu{\bf A}_\mu -
{\bf A}_\mu\times{\bf A}_\nu  ~.
\label{F}
\end{equation}

In the absence of the spin ordering, the Chern-Simons action
in terms of ${\bf A}_\mu$, has the following form (see the first term in
Eq.(3.1) of Ref.\cite{VolovikYakovenko}):
\begin{equation}
 S_{\rm CS}  = {N_3\over{64\pi}}\int
d^2x ~e^{\mu\nu\lambda}
\left({1\over 3}{\bf A}_\mu\cdot ({\bf A}_\nu \times
{\bf A}_\lambda)
+ {\bf A}_\mu\cdot {\bf F}_{\nu\lambda}\right)~.
\label{CS}
\end{equation}
Here $N_3$ is given by Eq.(\ref{3DTopInvariant}).

The chiral spin current is obtained as response to auxiliary
components ${\bf A}_i$ of the $SU(2)$ gauge field in the limit ${\bf A}_i
\rightarrow 0$ while  the physical field ${\bf A}_0$ is retained:
\begin{equation}
{\bf j}_i= \left({\delta S_{\rm CS} \over \delta {\bf
A}_i}\right)_{{\bf A}_i=0}= {N_3\over 16\pi } e_{ik} {\bf F}_{0k}
= {\gamma N_3\over 16\pi } e_{ik} {\partial {\bf H}\over \partial
x_{k}}~,
\label{SpinCurrent}
\end{equation}
The Eq.(\ref{SpinCurrent}) means that the spin
conductance is quantized in terms of elementary quantum
$\hbar/8\pi$ ($N_3$ is even number in these systems):
\begin{equation}
\sigma^s_{xy}=N_3{\hbar\over 16\pi } ~ .
\label{Quantization}
\end{equation}

For singlet superconductors there is no spin ordering, thus the
Eq.(\ref{Quantization}) is applicable. For the $d_{x^2-y^2}\pm i
d_{xy}$ superconductor the topological invariant
per spin projection is
$N_3=\pm 4$ (Sec. \ref{NontrivialMomentum}), and the
Eq.(\ref{Quantization}) reproduces the Eq.(21) of Ref.\cite{Senthill}.

For the triplet superconductor in the
$\hat {\bf z} (p_x + ip_y)$ state, spins in the
longitudinal direction (along  $\hat{\bf
d}=\hat{\bf z}$) remain  disordered: there is still $SO(2)$ symmetry
of spin rotations about the  $\hat{\bf z}$ axis in the
superconducting state under consideration. That is why the
Eq.(\ref{CS}) is valid if all auxiliary gauge fields
${\bf A}_\mu$ are chosen to be paralllel to the vector $\hat{\bf
d}=\hat{\bf z}$. Then  the
Eq.(\ref{SpinCurrent}) for the longitudinal spin remains valid:
\begin{equation}
j^z= {N_3\over 16\pi} e_{ik} F^z_{0k}
= {\gamma N_3\over 16\pi } e_{ik} {\partial H^z\over \partial
x_{k}}~,
\label{SpinCurrent2}
\end{equation}
Thus in triplet superconductors the spin Hall conductance for the
longitudinal spin is quantized according to Eq.(\ref{Quantization}):
\begin{equation}
\sigma^{s\parallel}_{xy}=N_3{\hbar\over 16\pi } ~ ,
\label{Quantization2}
\end{equation}
where for $p$ wave superconductor one has $N_3=\pm 2$ per layer.

\section{Topology of fermion zero modes on vortices.}
\label{CombinedTopologySspectrumVortex}

\subsection{Combined $({\bf r},{\bf p})$-topology in the vortex core.}
\label{InterplayTopologyCore}

As in the cases of the universality classes in the 3+1 and 2+1 systems,
the 1+1 system of electrons living in the vortex core also have
universality classes which determine the quasiparticle spectrum in the
low energy corner.  The fermions in this 1D ``Fermi liquid''
are chiral: the positive energy fermions have a definite sign of the angular
momentum $L_z$. In general case of arbitrary winding number
$m$, the number of fermion zero
modes, i.e. the number of branches crossing zero level as a function of
$L_z$, equals $-2m$ (see Ref.\cite{CallanHarveyEffect} and below). This
represents an analogue of the index theorem known for cosmic strings in
the relativistic quantum field theory (see
Ref.\cite{GeneralizedIndexTheorem} and references therein). The
difference is that in strings the  spectrum of relativistic fermions
crosses zero energy as a function of $p_z$, and the index
theorem discriminates between left-moving and right-moving
fermions, while in condensed matter vortices the index theorem  discriminates
between cw and ccw rotating fermions.
The index theorem follows from the interplay of ${\bf r}$-space and
${\bf p}$-space topologies.

After diagonalization over spin indices one finds that for each of two spin
components the Bogoliubov\,--\,Nambu Hamiltonian for
quasiparticles in the presence of the vortex has the form
\begin{equation}
{\cal H}=\left(\matrix{M(p) &
\Delta(\rho)e^{im\phi} \left({p_x +ip_y\over p_F}\right)^{N_3}\cr
\Delta(\rho)e^{-im\phi} \left({p_x -ip_y\over
p_F}\right)^{N_3}&-M(p)\cr}\right)~.
 \label{MicroHamiltonian1}
\end{equation}
Here $M(p)=p^2/2m^* - \mu$; $z$, $\rho$ and $\phi$ are
cylindrical coordinates with $z$ along the vortex axis; the order
parameter in the vortex has a form
$\Delta(\rho,\phi)=\Delta(\rho)e^{im\phi}$. The integral index $N_3$ is
related with the momentum space topology of superconductor: $N_3=0$ in
conventional
$s$-wave superfluid/superconductor;  the
$p$-wave superfluid $^3$He-A and chiral superconductors have
$N_3=1$ per each spin component; and in $d$-wave superconductors with
$d_{x^2-y^2}+id_{xy}$ order parameter one has $N_3=2$.

Introducing the angle $\theta$ in the $(p_x,p_y)$ plane in momentum space
the Eq.(\ref{MicroHamiltonian1}) can be rewritten in the form
\begin{equation}
{\cal H}=\left(\matrix{M(p) &
\Delta(\rho)\left({p_\perp\over p_F}\right)^{N_3} e^{i(m\phi +
N_3\theta)}\cr
\Delta(\rho)\left({p_\perp\over p_F}\right)^{N_3}e^{-i(m\phi +
N_3\theta)}&-M(p)\cr}\right)~.
 \label{InterplayTopologies}
\end{equation}
This form emphasizes the interplay between the  ${\bf r}$-space and
${\bf p}$-space topologies. The phase of the nondiagonal matrix element
contains simultaneously the winding in ${\bf r}$-space and the winding in
${\bf p}$-space.

Singularities in the Green's function caused by the vortex form the
manifold in the $({\bf r},{\bf p})$-space, which has non-trivial topology
in this extended space. Let us consider the manifold of zeroes of the
energy spectrum ${\cal H}^2=E^2({\bf r},{\bf p})$ in the classical
approximation, when both ${\bf r}$ and  ${\bf p}$ are $c$-numbers.
We use the simple case of the vortex with winding number $m$ in
conventional $s$-wave superconductor with $N_3=0$. While in homogeneous
state this superconductor is fully gapped, in the presence of the vortex
the quasiparticle energy spectrum
$E(x,y,z,{\bf p})$ becomes zero on a three dimensional manifold in 6D
space $(x,y,z,p_x,p_y,p_z)$. The equations for this 3D manifold are:
$p=p_F$ (where $M(p)=0$) and $x=y=0$: so it is the product of the 2D
Fermi surface and 1D line along the vortex axis $z$ . Each point of this
manifold is described by the same topological invariant,
Eq.(\ref{TopInvariant}),  as the topologically nontrivial Fermi point. If
instead of the 4-momentum
$(p_0,p_x,p_y,p_z)$ in Eq.(\ref{TopInvariant}) one introduces the mixed
coordinates
$(p_0,M,x,y)$ for given $z$, one obtains for the topological invariant in
Eq.(\ref{TopInvariant}) the value $N_3=2m$ (the factor 2 comes from two
spin projections).

Let us compare this manifold of zeroes within the vortex with the
manifold of zeroes in the homogeneous state of $^3$He-A. The
quasiparticle energy spectrum
$E^2(x,y,z,{\bf p})=(p^2/2m^*-\mu)^2 +c^2(p_x^2+p_y^2)$ of uniform
$^3$He-A with $\hat{\bf l}$ along $z$,  is zero at the Fermi
points at
$p_x=p_y=0$, $p_z=\pm p_F$ for arbitrary $x$, $y$ and $z$. In the combined
6D space $({\bf r},{\bf p})$ each Fermi point represents the
three dimensional manifold -- the $(x,y,z)$ space -- and each point of
this ${\bf r}$-space is described by the topological invariant $N_3=\pm
2$.

Thus we found that the gapless homogeneous state of
$^3$He-A and the conventional vortex with $m=1$ in the fully gapped
superconductor are described by the same topology in the 6D $({\bf r},{\bf
p})$-space. These two systems correspond to different orientations of
the 3D manifold of zeroes in this combined 6D space.  This implies
that these two states can be obtained from each other by continuous
deformation of the order parameter\cite{VolovikMineev1982}. This is the
reason why the core of the $^3$He-B vortex consists of the nonsingular
$^3$He-A (see detailed discussion in Ref. \cite{SalomaaRevMod}).

\subsection{Parity from combined topology.}
\label{ParityCombinedTopology.}

In the quasiclassical approach it is assumed that the characteristic
size
$\xi$ of the vortex core is much larger than the quasiparticle wave
length:
$\xi p_F\gg
1$, which works well when $\Delta\ll \mu$. In this quasiclassical limit
the description in terms of quasiparticle trajectories is
most relevant. The trajectories are almost the straight lines. The low
energy trajectories are characterized by the momentum ${\bf q}$ of the
incident quasiparticle on the Fermi surface, i.e.~with  $|{\bf
q}|=p_F$, and
the impact parameter $b$. Let us consider for simplicity the 2D or
layered
superconductors, so that ${\bf q}=p_F(\hat {\bf x}\cos\theta +
\hat {\bf
y}\sin\theta)$. Then substituting $\Psi\rightarrow e^{i{\bf
q}\cdot{\bf
r}}\Psi$ and ${\bf p}\rightarrow {\bf q}-i{\bf \nabla}$, and
expanding in
small ${\bf \nabla}$, one obtains the quasiclassical Hamiltonian for
the
fixed trajectory ${\bf q},b$:
\begin{equation}
{\cal H}=-i\tau_3{\bf v}_F\cdot{\bf \nabla}  +
 \Delta(\rho)\left(\tau_1     \cos(N\theta + m\phi)  -\tau_2
\sin(N\theta  + m\phi)\right)~,~{\bf v}_F={{\bf q}\over m^*}~.
 \label{QuasiclassicalHamiltonian}
\end{equation}

Since the spatial derivative is along the trajectory
it is convinient to choose the
coordinate system: $s=\rho\cos (\phi-\theta)$ -- the coordinate
along the trajectory, and
$b=\rho\sin (\phi-\theta)$ (see e.g. \cite{KopninSalomaa}). In this
system the Hamiltonian is
\begin{eqnarray}
 {\cal H} = -i{v}_F\tau_3 \partial_s +
 \tau_1  \Delta(\rho)   \cos \left(m \tilde\phi +(m+N_3)\theta\right)-
\nonumber\\
-\tau_2    \Delta(\rho)
  \sin \left(m \tilde\phi +(m+N_3)\theta \right)  ~,~~~\tilde\phi=\phi
-\theta~.
 \label{QuasiclassicalHamiltonian2}
\end{eqnarray}
The dependence of the Hamiltonian on
the direction $\theta$ of the trajectory can be removed by the
following
transformation:
\begin{eqnarray}
\Psi=
e^{i (m+N_3)\tau_3 \theta/2}\tilde\Psi,
\label{TransformationFunction}
\end{eqnarray}
$$\tilde{\cal H}=e^{-i (m+N_3)\tau_3 \theta/2}{\cal H}e^{i (m+N_3)\tau_3
\theta/2}=
  -i{v}_F\tau_3 \partial_s +$$
\begin{eqnarray}
+\Delta(\sqrt{s^2+b^2})\left(\tau_1  \cos  m
\tilde\phi  -\tau_2    \sin  m \tilde\phi\right),
\label{QuasiclassicalHamiltonian3}
\end{eqnarray}
\begin{eqnarray}
\tan
\tilde\phi={b\over s} .
 \label{TransformationHamiltonian}
\end{eqnarray}
Now $\theta$ enters only the boundary
condition for the wave function, which according to
Eq.(\ref{TransformationFunction}) is
\begin{equation}
\tilde\Psi(\theta +2\pi)=(-1)^{m+N_3}\tilde\Psi(\theta)
 \label{BoundaryCondition}
\end{equation}
With respect to this boundary condition, there are two classes of
vortices, which are determined by the parity
\begin{equation}
W=e^{\pi i(N_3+m)}~.
 \label{Parity}
\end{equation}
This parity, which comes from the combination of the ${\bf r}$-space
topology of the vortex (the winding number $m$) and the ${\bf
p}$-space topology of the ground state of superconductor, is
instrumental for the quantum spectrum of fermions in the core. Let us
consider vortices with
$m=\pm 1$.

\subsection{Quasiclassical fermion zero modes.}
\label{QuasiclassicalFermionZeroModes}

In quasiclassical approximation the quasiparticle state with the lowest
energy corresponds to such trajectories, which cross the center of the
vortex, i.e. with the impact parameter $b=0$. Along this trajectory one
has
$\sin\tilde\phi=0$ and     $\cos   \tilde\phi = {\rm sign}~ s$. So that
the Eq.(\ref{QuasiclassicalHamiltonian3}) becomes
\begin{equation}
 \tilde{\cal H}_{{\bf q},b} = -i{v}_F\check  \tau_3 \partial_s +
 \check  \tau_1  \Delta(|s|)  {\rm sign}~ s ~.
 \label{SupersymmetriclHamiltonian}
\end{equation}
This Hamiltonian is supersymmetric and thus contains the eigenstate with zero
energy. Let us write the corresponding eigen function  including all
the transformations:
\begin{eqnarray}
 \chi_{\theta, b=0}(s) =  e^{ip_F s} e^{i (m+N_3)\check  \tau_3
\theta/2} \left(\matrix{1 \cr -i\cr}\right)
\chi_0(s)~~,
\label{SupersymmetriclSolution1}
\\
\chi_0(s)=\exp{\left(-\int^s ds' {\Delta(|s'|)\over v_F} {\rm sign}~
s' \right)}~.
 \label{SupersymmetriclSolution2}
\end{eqnarray}

Now we can consider the case of nonzero impact parameter.  When $b$ is small
the third term in Eq.(\ref{QuasiclassicalHamiltonian3}) can
be considered as perturbation and its average over the wave function in
Eq.(\ref{SupersymmetriclSolution1}) gives the energy levels in terms of
$b$ and thus in terms of the continuous angular momentum $L_z=p_Fb$:
\begin{equation}
E(L_z,\theta)=-mL_z \omega_0~,~\omega_0  = {\int_0^\infty d\rho
 {\Delta(\rho)\over p_F\rho}\exp{\left(-{2\over v_F}\int_0^\rho d\rho'
\Delta(\rho')\right)}\over
\int_{0}^\infty d\rho \exp{\left(-{2\over v_F}\int_0^\rho d\rho'
\Delta(\rho')\right)}}~.
 \label{QuasiclasicalEnergy}
\end{equation}
This is the anomalous branch of chiral fermions which crosses zero energy in
semiclassical approximation, when the angular momentum $L_z$ is
consisdered as continuous variable. The fermions on this
branch rotate cw in the vortex with $m=-1$ and ccw in the vortex with
$m=+1$.

If one takes into account the discrete nature of the angular momentum
$L_z$, one obtains that the energy spectrum is discrete with the distance
between the levels  $\omega_0$ (which is called minigap) being of order
$ \Delta/(p_FR)$ where
$R$ is the radius of core. Typically $R$ is of order  coherence length
$\xi=v_F/\Delta$ and the minigap is  $\omega_0\sim \Delta^2/\mu\ll
\Delta$. The smallness of the distance between the levels justifies
the assumption that   $L_z$ can be consisdered as continuous
variable in a large temperature region
$\Delta^2/\mu\ll T \ll
\Delta$. In this energy range these bound fermionc states are the fermion
zero modes.

Let us consider now the energy scale of order or below the minigap.

\subsection{Quantum fermion zero modes and $W$-parity.}
\label{QuantumFermionZeroModes}

In exact quantum mechanical problem the generalized angular momentum $L_z$
has discrete eigenvalues. To find quantized energy levels we take into
account that the two remaining degrees of freedom, the angle $\theta$ and
the momentum $L_z$, are canonically conjugated
variables \cite{Stone,KopninVolovik}. That is why the next step is the
quantization of motion in the $\theta,L_z$ plane which can be obtained from the
quasiclassical energy in Eq.(\ref{QuasiclasicalEnergy}), if
$L_z$ is considered as an operator $L_z=-i\partial_\theta$.    For the
axisymmetric vortex,  the Hamiltonian does not depend on $\theta$
\begin{equation}
H= im\omega_0\partial_\theta
 \label{HamiltonianTheta}
\end{equation}
and has the eigenfunctions $e^{-iE\theta/m\omega_0}$.
The boundary condition for these functions, the
Eq.(\ref{BoundaryCondition}), gives the quantized energy levels, which depend
on the $W$-parity:
\begin{eqnarray}
E(L_z)=-mL_z \omega_0~,
\label{SpectrumCore1}\\
L_z= n ~~,~~W=+1 ~~;
\label{SpectrumCore2}\\
L_z =  \left(n+{1\over 2}\right) ~~,~~W=-1~~,
\label{SpectrumCore3}
\end{eqnarray}
where $n$ is integer.

The spectrum in Eq.(\ref{SpectrumCore3}) for the fermionic
states within the core of $m=1$ vortex in $s$-wave superconductor has
been obtained in Ref. \cite{Caroli}. This spectrum is not exactly
the fermion zero mode: it contains the small minigap of order
$\Delta^2/\mu$. It can be considered as fermion zero modes only in
semiclassical approximation, when the minigap is neglected. On the
contrary, the spectrum in Eq.(\ref{SpectrumCore2}) is exactly zero at
$L_z=0$. It is the true fermion zero mode. The difference between the
semiclassical and true fermion zero modes come from the $W$-parity in
Eq.(\ref{Parity}), which combines the ${\bf r}$-space winding
number and the ${\bf p}$-space topological invariant $N_3$.

\section{Conclusion}

We discussed some examples of the physical applications of the combined
${\bf r}$-space and ${\bf p}$-space topologies. The most fundamental
properties of the strongly correlated systems are
determined by this combined topology. The topological quantum numbers are
robust to any deformations (if they are not big enough to trigger the
quantum transition to the state with different topology). That is why,
for the investigation of the  complicated strongly correlated systems,
in many cases it is  enough to find the simplest perturbative model
within the same topological class and investigate it.

\end{document}